\documentclass[journal]{IEEEtran}
\usepackage{algorithm, algorithmic}
\usepackage{graphicx, soul}
\usepackage{hyperref}
\usepackage{amssymb}
\usepackage{slashbox}
\usepackage{epsfig}
\usepackage{amsmath}
\usepackage{xcolor}

\begin{document}
\title{Thinking Out of the Blocks: Holochain for Distributed Security in IoT Healthcare}

\author{Shakila Zaman, 
        Muhammad R. A. Khandaker,~\IEEEmembership{Senior Member,~IEEE,}
        Risala T. Khan,~\IEEEmembership{Senior Member,~IEEE,} 
        Faisal Tariq,~\IEEEmembership{Senior Member,~IEEE,}
        and Kai-Kit Wong,~\IEEEmembership{Fellow,~IEEE}

\thanks{Shakila Zaman is with the Department of Computer Science and Engineering, BRAC University, Dhaka, Bangladesh.}
\thanks{Muhammad Khandaker is with the School of Engineering and Physical Sciences, Heriot-Watt University, Edinburgh EH14 4AS, UK.}
\thanks{R. T. Khan is with the Institute of Information Technology, Jahangirnagar University, Dhaka 1342, Bangladesh.}
\thanks{Faisal Tariq is with the James Watt School of Engineering, University of Glasgow, UK}
\thanks{Kai-Kit Wong is with the Department of Electronic and Electrical Engineering, University College London, London WC1E7JE, UK.}
\thanks{\texttt{Submitted Version. Copyright may be transferred to IEEE anytime without prior notice.}}
}

%
%

\markboth{IEEE Internet of Things Journal,~Vol.~XX, No.~X, 20XX}%
{Holochain-Based Distributed Security}



\maketitle

\begin{abstract}
The Internet-of-Things (IoT) is an emerging and cognitive technology which connects a massive number of smart physical devices with virtual objects operating in diverse platforms through the internet. IoT is increasingly being implemented in distributed settings, making footprints in almost every sector of our life. 
Unfortunately, for healthcare systems, the entities connected to the IoT networks are exposed to an unprecedented level of security threats. Relying on a huge volume of sensitive and personal data, IoT healthcare systems are facing unique challenges in protecting data security and privacy. Although blockchain has posed to be the solution in this scenario thanks to its inherent distributed ledger technology (DLT), it suffers from major setbacks of increasing storage and computation requirements with the network size. This paper proposes a holochain-based security and privacy-preserving framework for IoT healthcare systems that overcomes these challenges and is particularly suited for resource constrained IoT scenarios. The performance and thorough security analyses demonstrate that a holochain-based IoT healthcare system is significantly better compared to blockchain and other existing systems.
\end{abstract}


%
\IEEEpeerreviewmaketitle

\section{Introduction}


The Internet-of-Things (IoT) is an exponentially increasing network of physical devices (the `things') that contain various embedded sensing, processing and communication technologies to collect and communicate sensory data through the internet \cite{SMRiaz15, jrnl_lwc_survey}. All interconnected entities of IoT networks are responsible to collect, store, process and exchange information with each other. With the amelioration of heterogeneous technologies, IoT is rapidly proliferating in all aspects of our life including smart healthcare, smart home, smart cities, agriculture, education, food industries, and many many more. In particular, the introduction of IoT applications in healthcare has the potential to revolutionize the sector where all the stakeholders will be interconnected to enable pervasive and universal healthcare for all regardless of their locations \cite{SMRiaz15}.


 The integrated connectivity amongst various entities of a healthcare system along with accurate and timely operations means that a massive amount of sensitive data will be shared with instant accessibility. A characteristic of an IoT-based healthcare network is that the data is originated at geographically distributed locations. Thus, the data is particularly vulnerable to unauthorized access and other malicious activities.  


One increasingly straining challenge for healthcare systems in both developed and developing worlds is the rapid expansion of aged population whose care requirement is different in nature and more demanding compared to the young population \cite{SBaker17Access}. The predominantly traditional physical/manual management of care system of aged population is further complicating the problem.  Furthermore, devices with very limited communication and networking capability and limited agility are exacerbating the problem. However, recent advances in flexible electronics \cite{wang2017flexible} and nano-bio sensors \cite{singh2018quantum} have the potential to address the critical healthcare problem mentioned above which was unthinkable even just a decade ago. Also, rapid progress in ubiquitous connectivity and networking solutions offered by 5G and emerging 6G systems will  enable remote healthcare management anywhere and round the clock \cite{jrnl_6g}. Progress in soft robotics for medical applications as well as medical informatics coupled with Immersive and eXtend Reality (IXR) will realize the dream of remote surgery \cite{jrnl_6g}. It will enable surgeons with certain expertise to assist and supervise robots to carry out the procedure from anywhere in the world provided that the critical latency and reliability requirements of the end-to-end connections are met.

Despite these promising development and innovation, data privacy and security in such a gigantic and  distributed network remains a major bottleneck for widespread implementation of smart healthcare systems \cite{Farahani2020,laplante2016internet}. Unless innovative solutions for security and privacy are designed and implemented, the smart healthcare system will remain vulnerable. This is evident from the ever increasing reports of numerous sophisticated cyber attack  on healthcare systems globally resulting in loss of sensitive health records as well as in significant downtime of the healthcare infrastructure \cite{ghafur2019challenges}. 


IoT based healthcare technologies offer numerous advantages including constant patient monitoring at a low cost, less error and significant saving in time. It also enables authorized doctors, staff and other technicians to access patient information online and real-time which improves the efficiency of the service significantly. As healthcare deals with the dynamic and real-time data such as patients' health status, prescriptions, test results, diagnosis, medical images and staff information, it is vitally important to keep all information extremely secure while allowing the right level of accessibility. With the acceleration of smart healthcare  functionalities, intruders can impede the quality of services in various ways such as slowing down normal operations or even bringing the infrastructure to standstill, injecting malicious data to alter critical information as well as tampering  medical devices to modify or take control patient records. Moreover, IoT devices are lightweight  and resource-constrained with limited memory, low computation power, and limited energy supply. Therefore, providing security and privacy using traditional cryptographic approaches are impractical in most of the scenarios and quite challenging to implement \cite{jrnl_lwc_survey, jrnl_secrecy}. 



As IoT healthcare devices have resource constraints, conventional cryptography techniques such as advanced encryption standard (AES) and Rivest–Shamir–Adleman (RSA) are not suitable for securing massive amount of sensitive information. Therefore, lightweight cryptographic algorithms such as \emph{lightweight SIMON} ciphers are employed in IoT healthcare applications, which offers reduced time complexity and pragmatic trade-off between security and services \cite{alassaf_enhancing_2019}. 
However, due to the heterogeneity and dynamic environment of IoT healthcare technologies, attackers can still post various threats that make the system vulnerable to data theft and tampering. Machine learning (ML) and deep learning (DL) based security approaches have gained popularity as they possess advanced capability of tackling security challenges. In \cite{newaz_healthguard_2019}, an ML based HealthGuard framework was proposed which combined artificial neural network (ANN), decision tree, random forest and K-nearest neighbor algorithms to detect various malicious activities. An ML-enabled biometric security method was also introduced to train electrocardiogram (ECG) signals to authenticate users of a medical system \cite{SPirbhulal_2019}. 


Despite the progresses mentioned above, an IoT healthcare system remains vulnerable as most of the cryptographic solutions are centralized which are incompatible to the distributed and heterogeneous nature of the IoT healthcare system and present the risk of a single point failure. They also suffer from inability to tackle complex attacks in resource-constrained IoT networks. Additionally, the performance of ML and DL algorithms is highly dependant on high-quality sample data for training which is not always readily available. Consequently, real-time distributed security approaches such as blockchain-based approaches have recently emerged as a promising alternative to provide effective responses to various privacy, security and authentication challenges in IoT healthcare systems. In principle, blockchain is a distributed ledger technology (DLT) that stores information in a series of integrated blocks which are difficult to tamper. Due to its transparency and fairness, blockchain has a wide range of  applications including real-time IoT operating system \cite{BC_2018}, secure smart home \cite{smarthome_2017}, personal identity, supply chain management \cite{cole_blockchain_2019}, real-estate processing platform \cite{realstate_2019} as well as smart healthcare \cite{dwivedi_decentralized_2019}, to name a few. The most popular application of blockchain is the cryptocurrency application such as bitcoin which is expected to save money and time of business entities.

Blockchain is also well known to provide distributed security for data in healthcare environments. Blockchain-based security solutions have been proposed in 
\cite{griggs_healthcare_2018} for remote healthcare operations where the patients are entities of a body area network and collect various medical data to share with authorized entities through an overlay network. Blockchain technique also provides a distributed framework to store large-scale medical information in the clouds, ensures authorized access on the database, guarantees the integrity of each modification and confirms secure transactions among blockchain entities through consensus algorithms 
\cite{JXu_2019}.  


As each entity of a blockchain network stores all users' transactions in a chain, the memory requirements at blockchain entities escalate with increasingly longer chains thereby jeopardizing practical applications of resource-constrained IoT devices. Another consequence of increasing transactions in a longer chain is that it requires a very large bandwidth and data sharing giving rise to security vulnerability. The challenge is further exacerbated by the requirement of additional computational energy for mining and consensus algorithms \cite{mcghin_blockchain_2019}. To validate any transaction in a blockchain, all the nodes will start mining and only the first node who is successful in the mining process will be allowed to validate the transaction. From the aspect of computational time, this is a complete wastage of time for the rest of the nodes who attempted the mining process but were unsuccessful. Hence, blockchain gives rise to redundant computational overheads for resource-constrained IoT healthcare systems. Therefore, a practically feasible solution is required for IoT systems that can overcome this challenge. It is vitally important to ensure that the DLT is implemented in an inherently secure and privacy-preserving manner in the distributed IoT healthcare setting, while being less complex and less resource-hungry.


Holochain is an emerging technology that provides an open source distributed network infrastructure to communicate securely without inheriting the huge storage and data exchange requirements like blockchain \cite{brock_holochain_2020}. Holochain magically performs the task by combining two underlying techniques: (i) distributed hash table (DHT) and (ii) hash chain. DHT is focused on data propagation issues and hash chains are built to preserve data integrity \cite{janjua_proactive_2020}. One of the main visions of holochain is to reduce dominant characteristics of certain network entity. For instance, most of the applications are based on the client server model which normally imposes restrictions on resource utilization. Contrary to this model, holochain aims to build a completely distributed network. DHT replaces the need of centralized control of flow or management of data. DHT can be implemented and utilized in IoT healthcare networks for storing the chain of transition data in each individual node to ensure the autonomous nature of a holochain-based network. The DHT concept can be utilized to share data with each other and  provide an actual distribution framework. The most significant aspect of storing data in DHT is that the  network will not become congested like the blockchain-based network does. The DHT of holochain allows the network to provide scalable performance. To summarize, all these attributes make holochain an attractive candidate for IoT healthcare systems.

In this paper, we propose a holochain-based IoT healthcare framework that mitigates the security and privacy challenges and offers a low-complexity, highly-secure alternative to blockchain. Our main contributions are listed below:
\begin{itemize}
    \item A holistic holochain approach to address the security and privacy issues in IoT healthcare systems;
    \item Critical analysis with regards to the superiority of the holochain framework over blockchain based systems;
    \item Systematic algorithms design for holochain implementation as well as validation and authentication procedures;
    \item Security performance analysis of the holochain-based IoT healthcare network as compared to blockchain and other classical cryptographic systems.
    \item Discussions on the challenges of the implementation of holochain based IoT healthcare systems followed by comprehensive discussion on future research directions.
\end{itemize}

The rest of this article is organized as follows. Section II describes the vulnerabilities of IoT healthcare systems. Section III introduces the role and working principle of blockchain in IoT security in general. Section~\ref{sec_adv} then provides a comprehensive discussion on how holochain can address the limitations of blockchain. In Section V, a thorough security performance analysis of existing security mechanisms is given. The proposed holochain-based IoT healthcare framework is presented in Section VI followed by a component-wise discussion in Section VII and its implementation in Section VIII. A security analysis of the holochain-based IoT healthcare network is provided in Section IX. Finally, some key challenges and promising future research directions are provided in Section X before we conclude the article in Section~\ref{sec_con}.

\begin{figure*}[htp!]
\centering
\includegraphics[width=0.8\textwidth]{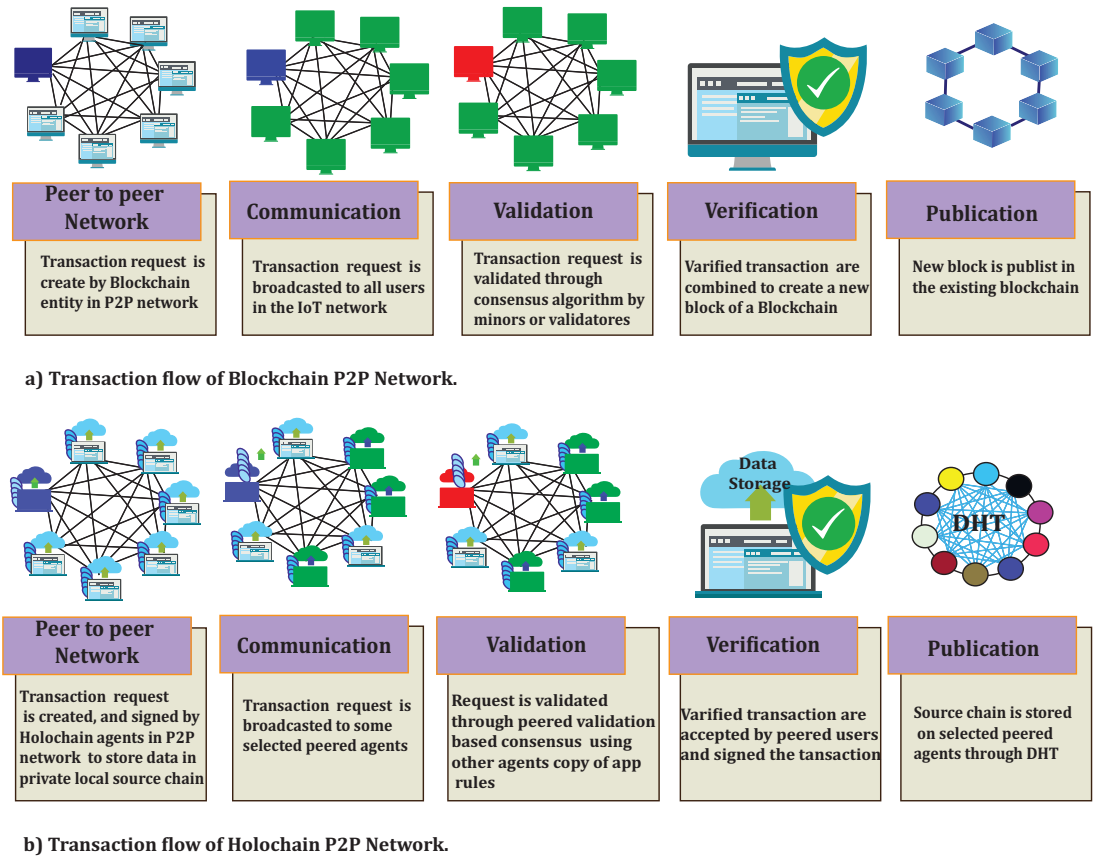}
\caption{The transaction validation processes of blockchain and holochain.}\label{fig:BvsH}
\end{figure*}

\section{Vulnerability of IoT Healthcare Systems}

Cyber-attacks pose a serious threat to our daily life as we are increasingly relying on IoT connections in all aspects of our life which potentially exposes our personal data, financial information, computers and other devices, home and work and even medical data to hackers and malicious entities. Anything connected to the IoT network has the risk of being hacked or compromised. From a pacemaker in heart to our infotainment in cars, everything is connected to the IoT network and thus is exposed to the threat of cyber-attacks. When it comes to healthcare, it is even more alarming. 
For example, a CT lung scan can show the ravaging signs of COVID-19 and the inflammatory response, the cytokine storm. What if the CT scan was wrong? An attacker with access to the medical imagery can alter the contents to cause a misdiagnosis. The attacker can even add or remove evidence of some medical conditions, e.g., inject/remove lung cancer from a scan, add/remove evidence of aneurysms, heart disease, blood clots, infections, arthritis, cartilage problems, torn ligaments or tendons, tumours in the brain, heart, or spine, and other cancers \cite{ct_gan}. These are no longer science fictions but are happening in real life nowadays. 

In 2018, a malicious attack was designed to hack hospital CT scans, generating false lung tumours that conformed to a patient’s unique anatomy, leading to a misdiagnosis rate in excess of $90\%$ \cite{ct_gan}. Furthermore, researchers at Harvard University tested adversarial attacks against algorithms used to diagnose skin cancer images, demonstrating that such attacks required only modifying a few pixels in the original biopsy picture to deceive a diagnosis \cite{Finlayson1287}. 
Marie Elisabeth Gaup Moe, a cybersecurity specialist, successfully hacked her own pacemaker to show how vulnerable we are to cyberattacks. 

The real risk is that a hacker could potentially take over the communication to the medical device; switch it off, make it malfunction, or falsify the information that is sent to the doctor. Researchers have also shown that it is possible to do something called a battery-draining attack on a pacemaker implant. Many recent cyber-attacks have shut down hospitals/healthcare systems and caused deaths. In 2017, the global ransomware attack, WannaCry, took hold across multiple continents and inflected over $600$ organizations including $34$  hospital trusts in the UK that were locked out of their digital systems and medical devices, such as MRI scanners \cite{wannacry}. WannaCry impacted patient care directly, costing the organization $\pounds 92$m ($\$116.4$m) and leading to $19,000$ cancelled appointments. These examples are just a sample of how AI can automate the manipulation of medical datasets, expanding a cyber attack’s impact through health and biotech industries.

While raising awareness at all levels is certainly necessary, a proactive, insightful, forward-thinking approach is desperately needed to make the necessary changes to protect individuals. 
This is so true of any solution built using the current ageing approach to software. The current coding approach is too detailed for the human mind to comprehend. Hence many gaps in the logic are missed, creating loopholes, which in turn, creates doorways o backdoors for hackers to exploit. 

In this article, we explore DLTs which are `secure-by-design' and implementable under resource-constrained IoT settings. Although this article focuses on medical applications facilitated through the IoT, the proposed \textit{secure-by-design} technology will have wider implications. For example, hacking of transport systems (including autonomous driverless cars) also has the potential to cause devastating accidents. Hacking of financial systems can cost billions from businesses, wipe out lifetime savings in pension schemes, and critical infrastructure could also be damaged, and the list goes on. Therefore, we need to include cyber-security in the design phase of our connected devices and should not see it as an add-on feature.  
We need to ensure that the system is secure by nature, which is not only profitable but also safe at the same time, such that people with severe health conditions and requiring special care can feel safe while being assisted with care facilities.


\section{Blockchain for IoT Security}
Blockchain is a DLT in which a chain of blocks allows authorized IoT users to store private or shared transactions in a decentralized fashion. Every node (IoT devices) involved in the transaction maintains a copy of each transaction blocks of the network. Each block consists two parts: a header and a body. The body of the block mainly stores the data or information of certain transactions. A header holds several items including the previous contents hash, timestamp, nonce (a solution to the cryptographic puzzle), target value (for mining) and Merkel root (all transactions root value of block) \cite{BC_2018}. 
The transaction flow of a block of a blockchain network is shown in Fig. \ref{fig:BvsH}. 

Before storing the linked blocks of transactions among various IoT nodes, blockchain utilizes the benefits of public-key cryptography to sign the transactions. In blockchain, once a block is entered in the cryptographically immutable chain, data can never be changed or removed. A blockchain stores every single transaction of a user in a specified time. It is linked automatically  with the previous block using the hash technique. Whenever a new block is created, mining is performed for validating the block by some selected miners (random IoT nodes) to solve a cryptographic puzzle called consensus algorithms. There are several variants of consensus algorithm such as Proof-of-Work (PoW), Practical Byzantine Fault Tolerance (PBFT), Proof-of-Stake (PoS) and Delegated Proof-of-Stake (DPoS) \cite{hammi_bubbles_2018}. The miner who first finds a nonce of the complex computational puzzle becomes the owner of the new block using its private key with timestamp and broadcasts the block to all the connected IoT nodes in the network to store in their local blockchains.  Thereafter, other nodes will accept the new block using a verification process.

To provide IoT security, hash and digital signatures of blockchain are exploited. In blockchain, hash is used during the storing process of every current block which consists of the hash value of the previous block as well as its own contents hash. The hash helps to provide a unique fingerprint as well as to ensure verification of the blocks to confirm that it has not been tampered or changed during transactions or storing process. The very first hash of a blockchain is formulated using the first block or genesis contents followed by the afterword calculation using the previous block's hash and its own hash. Therefore, the chain of the hash can be used to easily detect any tempering of the history of the block or transaction. For example, in a smart healthcare system, a block may hold patient's billing information which is hashed with a specific hash function known by authorized users only. If an unauthorised user tries to access or modify the information, the hash value of the original message will not be matched. Therefore the receiver can simply discard the message through the verification process and ensures data integrity of a transaction. On the other hand, digital signature is used to ensure the authenticity of an IoT user or the origin of the messages. Digital signature not only provides message authentication but also ensures non-repudiation of a message. As only the authorized sender has the knowledge of the digital signature key, the receiver can use the data with signature to provide authentic information and to refute any future disagreements with third party. 

Moreover, blockchain uses smart contracts which are self-executing predefined agreements maintained by the peers of the IoT networks to structure their relationships. A smart contract restricts the accessibility or functionality of the individual users, which is dependent on specific applications. The smart contracts provide an additional layer of security which is digitally signed by the user’s private key and can only be decrypted by the public key of the shared authorized peers.

Despite the above-mentioned benefits of blockchain, there are a number of challenges that make the blockchain-based system mostly impractical for large-scale deployment. As mentioned before, the complexity and resource requirements (both in terms of storage, processing and communications) increase rapidly with the increased IoT network size as the hash length of blockchain will grow with more transactions. Every new entity involved will need relatively larger memory space to store a longer hash. This will also inevitably require much more spectrum/bandwidth for distributed authentication. This is particularly challenging for healthcare systems as many of entities in the network are envisaged to be operating on energy and resource constraints. Therefore, a lightweight solution is of paramount need in order to enable a practically implementable secure IoT healthcare system.



\section{Holochain: A Lightweight and Scalable Security Solution for IoT Healthcare Networks}
 \label{sec_adv}

Holochain is an emerging technology that has the potential to address the drawbacks of the blockchain in enabling secure IoT healthcare networks. Holochain is based on an open source distributed network infrastructure that communicates securely without inheriting the huge storage and data exchange requirements of a blockchain. In this section, we demonstrate how holochain offers better solutions in contrast to blockchain for realizing secure IoT healthcare systems.

\subsection{Better Scalability}
Blockchain is a data-centric distributed security approach where the main objective is to create a single shared block of data among all authorized users in the network. Thus the size of the data will therefore increase with the number of network entities involved in each transaction and is not scalable \cite{Harris-Braun_Holochain_2019}. In contrast, each holochain application (hApp) is maintained by an agent that can independently participate in data encryption, storing transaction in a unique source chain of a holochain network and share the required data with a peered agent. This agent-centric holochain approach is highly scalable.

\subsection{Reduced Network Traffic}
Since holochain combines digital signature and DHT, it could be an effective alternative of blockchain to ameliorate the performance of retrieving information from a distributed peer-to-peer (P2P) network. Each agent in a holochain network stores its individual data locally. In IoT networks, many devices use offloading concepts, fog nodes, or clouds to store their database due to the limitation of memory and computational power. However, each agent is capable of computing their own hash value and shares a significant part with other peers using DHT. In contrast, all peers of a blockchain network store an indistinguishable copy of the transmission which requires significantly more communication exchanges between the nodes. Moreover, additional bandwidth is required by each entity which significantly increases the network bandwidth consumption and affects the scalability. However, in holochain, agents do not need to share their individual transaction information with all other peers of the network except some nodes which will have a backup whenever the owner goes to offline. Thus, holochain can significantly reduce the amount of bandwidth requirement and traffic in the network \cite{Frahat_2019}.

\subsection{Low-Complexity Transaction Validation}
In blockchain, miners are responsible for validating new transactions by solving a mathematical problem. Any network node can act as a miner and initiate mining anytime. For example, if there are 20 network nodes and 10 of them start to mine for validating a transaction. The node that finds the solution of the mathematical challenge earliest will validate the transaction. A miner can cooperate with the other nodes and mine simultaneously. The involvement of the other 9 nodes in the mining process are complete wastage of time and resource. If a network split occurs amid the mining process, then it becomes difficult to recognize which part of the network is still active resulting in new security issues for that transaction. This may turn out to be crucial in some situations such as during a payment-related transaction. For example, if a transaction holds cryptocurrency information, splitting events may trigger disagreements and uncertainty between the users \cite{islam_why_2019}.  

In contrast, holochain allows individual nodes to validate its own transaction and neighbour nodes with a predefined distance are allowed to do the secondary validation of that transaction when the transaction information is sent to them along with some other pre-settled information. As only few nodes keep the copy of the transaction rather then all the nodes in the network, the memory space and the amount of information exchange are significantly lower than blockchain. 

\begin{table*}[ht!]
    \centering
    \scriptsize
\caption{Difference between holochain and blockchain}\label{5g6g}
\begin{tabular}{ |c|c|c|c|} 
 \hline
{\bf Characteristics} & {\bf Blockchain} & {\bf Holochain} & {\bf Reference} \\ 
\hline\hline
Chain contains & Hole ledger data  &   only a part of entire ledger data &  \cite{Frahat_2019}  \\ 
\hline
Cryptocurrency & Bitcoin & Holo fuel &  \cite{Web_2020}, \cite{hostholofuel-model_2019}\\ 
\hline
Chain approach & global data stored in block & Data stored in individual agents &  \cite{Frahat_2019}, \cite{Harris-Braun_Holochain_2019} \\
\hline
Mining approach & Mining is essential  & No mining approach is required & \cite{Frahat_2019} \\ 
\hline
Consensus  & POW, POS, DPOS, PBFT & No consensus is required  &  \cite{Frahat_2019}, \cite{Holochain_org} \\ 
\hline

Processing Cost & High & Low  &  \cite{noauthor_redistributive} \\ 
\hline
Shared transition with & All nodes & Few nodes  & \cite{diojdescu_research_nodate} \\ 
\hline
Error Handling & Data rejection & Data rejection, Node blacklisting  &  \cite{Holochain_org}\\ 
\hline
Data redundancy  & Extreme &  Optimal  & \cite{Frahat_2019}\\ 

\hline
Scalability & Low & High  &  \cite{diojdescu_research_nodate} \\ 
\hline
Energy consumption  & High  & Low &  \cite{diojdescu_research_nodate} \\ 
 \hline

Transition degree & Maximum $1000$ per second & Maximum $56000$ bps/Hz &  \cite{diojdescu_research_nodate}\\ 
 \hline
Data integrity & Validation is ensured by Miners & Validation is ensured by previous cryptographic record &  \cite{Holochain_org}\\
\hline
\end{tabular}
\end{table*}

\subsection{Efficient Consensus Mechanism}

Unlike blockchain, holochain does not require a global consensus mechanism. Holochain is designed to provide autonomy for each user or a group of users who can validate the transaction without any global consensus. The validation processes of both blockchain and holochain based networks are illustrated in Fig.~\ref{fig:BvsH}. Evidently, holochain is more efficient than blockchain. To validate a transaction, blockchain sends the current transaction to all nodes for storing full node information whereas holochain requires only a few holo-hosts who are involved in running the same application to validate the current transaction without requiring a global consensus. Moreover, the validation process, data ownership rights and network governance are managed by agents and creators only. In some cases, it may happen that the data, which is posted to validate nodes or transactions, itself is not authorized or valid. To address this problem, \textit{hashed fingerprint} is used to help detect the authentication of a transaction.

 \subsection{Communication Resource Efficiency}
 
In a P2P network, blockchain relies on persistent communication among the distributed users. Moreover, it involves a set of miners to process and validate a transaction, and store it in all the users. The consensus mechanism which is a significant part of the blockchain, also demands a large number of communication channels that limit the transaction throughput of the network \cite{zhang_how_2021}. On the contrary, the holochain consensus mechanism is agent-centric and does not require frequent communication with other nodes that greatly reduces the number of occupied communication channels.    
 
\subsection{Operating Time and Memory Efficiency}
An inherent property of blockchain is to have the same transaction information in all nodes for providing data integrity throughout the hash tree. In many practical applications, a particular user's data may not be of interest to others, but a blockchain network enforces all users to store all of the information resulting in increased data processing time and larger memory space. Given that many IoT healthcare devices are lightweight, this is detrimental to their design objective. As a consequence, the entire system becomes slower compared to the holochain-based counterpart \cite{Frahat_2019}.  For example, in a smart healthcare management system, doctor X, and many others if not all, do not need to know the glucose level of patient Y. In blockchain, doctor X and others nodes are also imposed to store the transaction regarding glucose level information of patient Y. However, in holochain, only some selected agents will store it to ensure data integrity and store the transaction locally that saves memory as well as processing time. Moreover, hApp agents share the transaction data using DHT which requires less space and makes the network faster.
 
\begin{figure}[htp!]
\centering
\includegraphics[width=\linewidth]{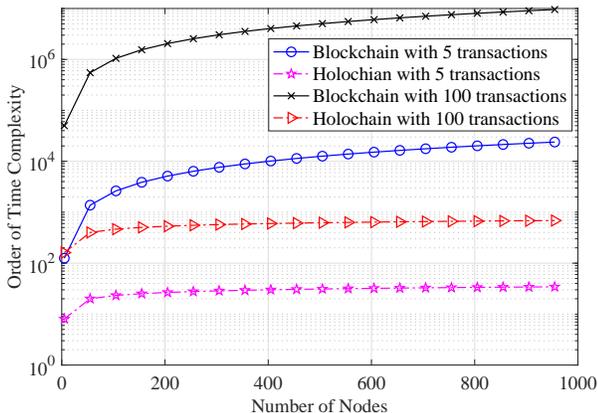}
\caption{Time complexity order of blockchain and holochain networks.}\label{fig:HL}
\end{figure}
 
\subsection{Efficiency in Large-Scale Networks}
Since blockchain technology monitors and stores all transactions at every node connected to the network, the network load increases rapidly with the increasing number of users leading to high inefficiency in large-scale networks. For instance, if a network consists of $100$ nodes, then the network efficiency will be reduced $100$ folds due to the increased data redundancy as well as time-complexity for each transaction. On the contrary, for holochain, the processing tasks only escalate linearly and distribute the processing loads among other nodes of the network. Considering the example above, if a holochain network includes $100$ agents, then the whole network load will be distributed among $100$ nodes and each node will only process a small fraction of the total transactions. Therefore, most nodes will save significant processing capacity. To generalize this, the average time-complexity of a blockchain network implementing a Bitcoin structure is given by \cite{Harris-Braun_Holochain_2019}
\begin{equation} \label{eq1}
\begin{split}
	\Omega_{Blockchain} \in	\mathcal{O} (n^2 * m), \nonumber
\end{split}
\end{equation}
where $n$ is the number of nodes and $m$ is the number of network-wide transactions required. In contrast, the average time-complexity for the holochain framework is given by
\begin{equation} \label{eq2}
\Omega_{Holochain} \in \mathcal{O} (m*(\log(n)+c)), \nonumber
\end{equation}
where $c$ is the application-specific complexity parameter \cite{Harris-Braun_Holochain_2019}. 

Exploiting the above time-complexity definitions, Fig.~\ref{fig:HL} presents a comparative analysis of the time-complexity order for both blockchain and holochain networks against the number of nodes. It can be observed that the order of time-complexity for a blockchain network increases exponentially with the number of nodes while the average order of time-complexity in a holochain network remains largely settled for a larger number of connected nodes.

\subsection{Better Protection Against Consensus Based Attacks}
As blockchain is a consensus-driven technique, a handful of attacks may target to disrupt the consensus operations. A huge number of nodes require to detect and prevent the attacks requiring high computational capacity. Conversely, the agents in holochain are mostly accountable for their own transaction history and consistently audit others' holo-currency to validate the credit spending status. Therefore, the agents only need to build trust in its own code and thus are less prone to consensus-based attacks like majority attacks, sybil attacks, PoW attacks, selective drop attacks, and etc.

\subsection{Application-Dependent Validation Function}
In blockchain, all the applications of the same network must go through the same validation rule to validate each transaction. However, not all the transactions have the same importance and deserve the same resources for validation. For example, in healthcare systems, the information of the appointment schedule of a doctor is not as important as the medical report or live patient monitoring data sent to the doctor. For a blockchain network, the same validation rule will be applied for both cases but in a holochain, information validation functions can be designed in a way that they will incur resource cost that are proportionate to their level of importance. Thus the complexity of validation can be adjusted to make the best use of the available resources while ensuring the required security, privacy and authenticity.

\subsection{Cost Effective Solution}
Since the computational cost and complexity of blockchain are significantly higher than those of holochain, particularly in large-network scenarios, holochain is a lot more efficient in terms of energy cost and cost of required equipment. 

In light of the above comparative analysis, it becomes apparent that holochain offers a more viable option compared to blockchain for IoT healthcare systems. Table \ref{5g6g} provides a comprehensive summary of the comparison between holochain and blockchain technologies and the associated references for further reading. It is evident that holochain is convincingly a better choice for distributed real-time systems. In the following sections, a novel holochain based IoT smart healthcare framework is proposed that exploits the benefits discussed above.


\section{Our Holochain-Based IoT Healthcare Model}

\begin{figure*}
\centering
  \includegraphics[width=.8\textwidth]{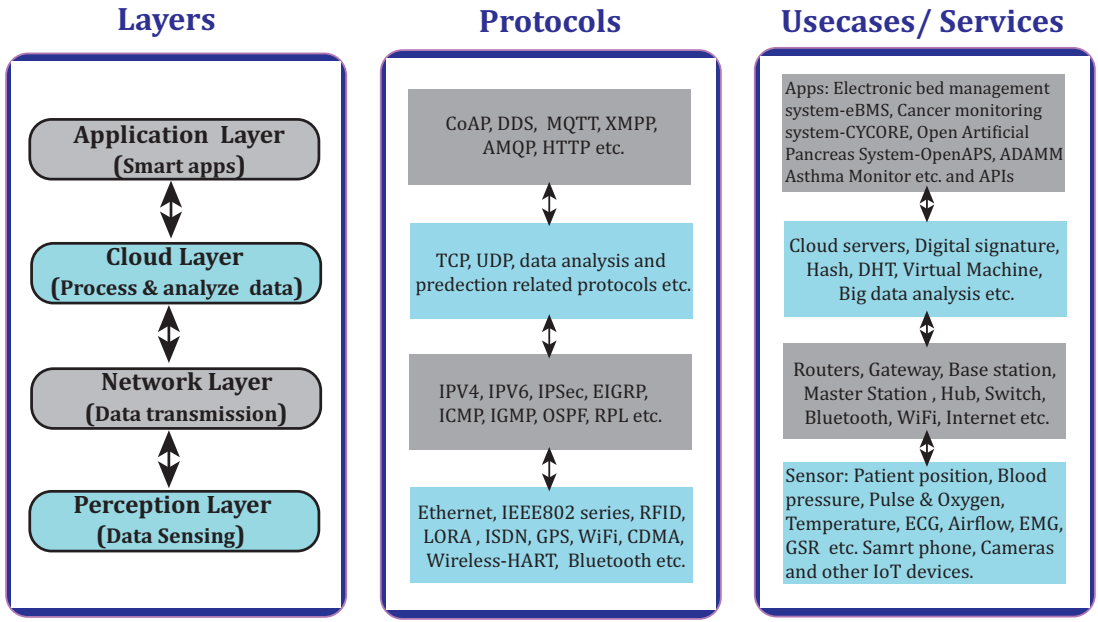}
\caption{Layer-wise protocols and technologies of the IoT healthcare architecture.}
\label{fig:Layer}       
\end{figure*}

In this section, a novel holochain-based smart IoT healthcare system is proposed which guarantees strict data integrity as well as high level of network security. The proposed IoT healthcare framework constitutes four main layers: (1) IoT (2) perception layer, (3) network layer, cloud or processing layer and (4) the application layer. Fig.~\ref{fig:Layer} shows the functionalities and protocols in each layer of the IoT healthcare system. The functionalities of each layer are discussed in details below.

\subsection{The Perception Layer}
Perception layer is responsible for sensing and collecting necessary information such as patient's health data. IoT nodes sense and collect the data and transfer them using various protocols and systems including Ethernet, IEEE802 series, wireless sensor network (WSN), global positioning system (GPS), wireless-HART and Bluetooth. A robust trust assessment system is used to collect data from authorized users only \cite{oliveira_mac_2019}, 
Fig.~\ref{fig:holochain} describes a holochain-based IoT healthcare framework that interconnects various medical entities such as patients, doctors, staff, technologists, pharmacists as well as medical devices in the perception layer. Every entity of the healthcare system can have multiple hApps. A unique set of logic-based rules are employed to provide the specific services using these hApps. For instance, a patient can use the QardioCore app that is an ECG monitoring system to deliver incessant health grade information \cite{bleda_quality_2019,ECG_2020}. The same patient can also utilize various other apps connected to smart IoT devices such as patient position sensors, blood pressure sensor, pulse sensor, oxygen level sensor and temperature sensor to collect and analyze various health related data for monitoring physical or mental conditions. The health related information collected from the perception layer can be transferred to a specialist for real-time monitoring through the internet.


\subsection{The Network Layer}
Network layer accepts the forwarded data from the perception layer which are processed information by various hApps for IP addressing. This layer ensures reliable transmission paths using various common protocols like Internet Protocol version 4 and 6 (IPV4, IPV6), Internet Protocol Security (IPSec), Enhanced Interior Gateway Routing Protocol (EIGRP), Internet Control Message Protocol (ICMP), Internet Group Management Protocol (IGMP), and Open Shortest Path First (OSPF), etc. \cite{Kharrufa_2019,rayes_internet_2017}. Network layer handles the transaction and provides services using heterogeneous devices and technologies including routers, gateway, base station, master station, hub, switch, Bluetooth and WiFi. After processing the packets, this layer transfers the trusted information to the upper layer known as the cloud layer which is responsible for managing the storage and sharing of the trusted values among the IoT nodes in a distributed manner.

\subsection{The Cloud Layer}
Since IoT devices are resource-constrained, sensitive medical information could be stored and preserved in the cloud, authorized parties (e.g., doctors, insurance providers, medical staff, pharmacy, etc.) can conveniently share the information with each other. Like patients, other entities can also store their information in the cloud and share that sensitive information with authorized peers for augmenting the performance of healthcare services. Transmission Control Protocol (TCP), User Datagram Protocol (UDP), and ML, data analysis and data predictive protocols are the common protocols in this layer \cite{rayes_internet_2017}. 
To ensure security and data integrity in the cloud layer, this work uses distributed holochain in cloud devices. 


\subsection{Application Layer}
The upper layer of the IoT network is responsible for information formatting and presentation. This layer defines a set of rules for transferring the message. Constrained Application Protocol (CoAP), Data Distribution Service (DDS), Message Queue Telemetry Transport (MQTT), Extensible Messaging and Presence Protocol (XMPP), Advanced Message Queuing Protocol (AMQP) and Hyper Text Transfer Protocol (HTTP) are the well-known protocols dedicated to the application layer 
\cite{seleznev_industrial_2019}. Application layer introduces a variety of healthcare services. For example, electronic bed management system (eBMS) is used for managing beds in the hospital. CYCORE is another crucial application which is used for cancer patient monitoring. Open artificial pancreas system (OpenAPS) is designed to help automate the insulin delivery system. To summarize, the application layer is responsible for delivering app-based services via direct communication with the users.

\section{Key Components}
The proposed framework consists of holochain, holofuel, hashchain and the DHT. The functionalities of all these major components along with the source chain structure of the holochain are described in detail here \cite{Holochain_org}.

\begin{figure*}
\centering
  \includegraphics[width=0.9\textwidth]{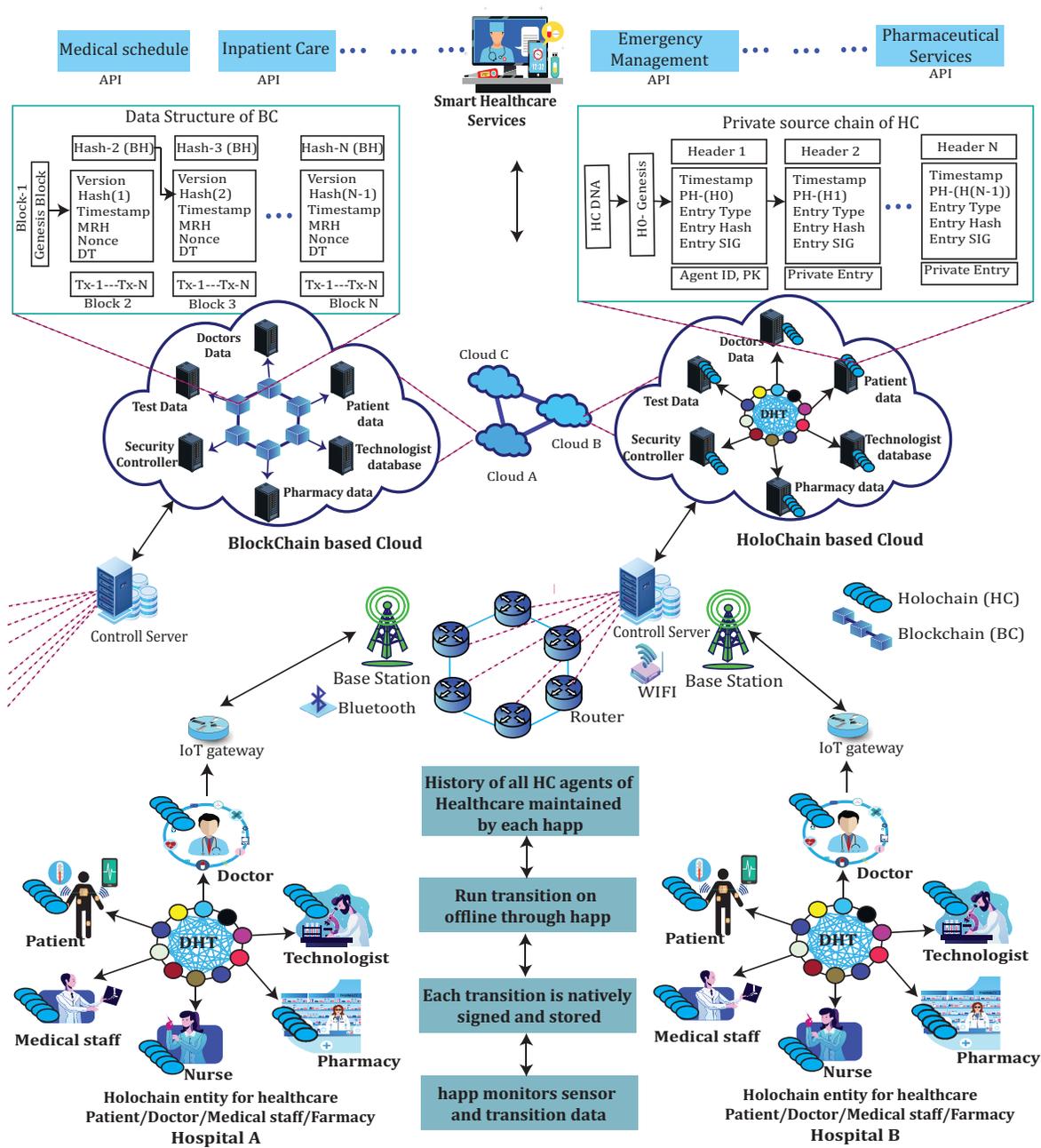}
\caption{The holochain-based IoT healthcare architecture.}
\label{fig:holochain}       
\end{figure*}

\subsection{Holochain} 
Holochain is a well-organized energy-efficient DLT for the next-generation internet that utilizes P2P network facilities for handling agent-centric commitment and consensus model among users. The fundamental benefit of a holochain network is to have an individual secure ledger that ensures individuality even when it is communicating with other peers of the network. Thus, it permits fully distributed computing.

\subsection{Holo Fuel}
Like blockchain-based Ethereum and Bitcoin, holochain also introduces an electronic currency to support the payment system of the holo-hosts, which is known as holo fuel \cite{noauthor_redistributive}. Holo fuel is also defined as a mutual-credit system which is required to perform millions of daily transaction for holo-hosts or users. Holochain network providers are accountable for managing and providing service regarding transaction fees through specific protocols. The price or value of the holo fuel is operated by the computing capacity of a host in the network. Moreover, cyptocurrency transaction and balance status are also signed, stored in an individual user. They are shared using DHT for validation like all other transactions. Two authorized agents can make a transaction only if they have a sufficient credit balance 
\cite{hostholofuel-model_2019}.

 \begin{figure*}
\centering
  \includegraphics[width=.7\textwidth]{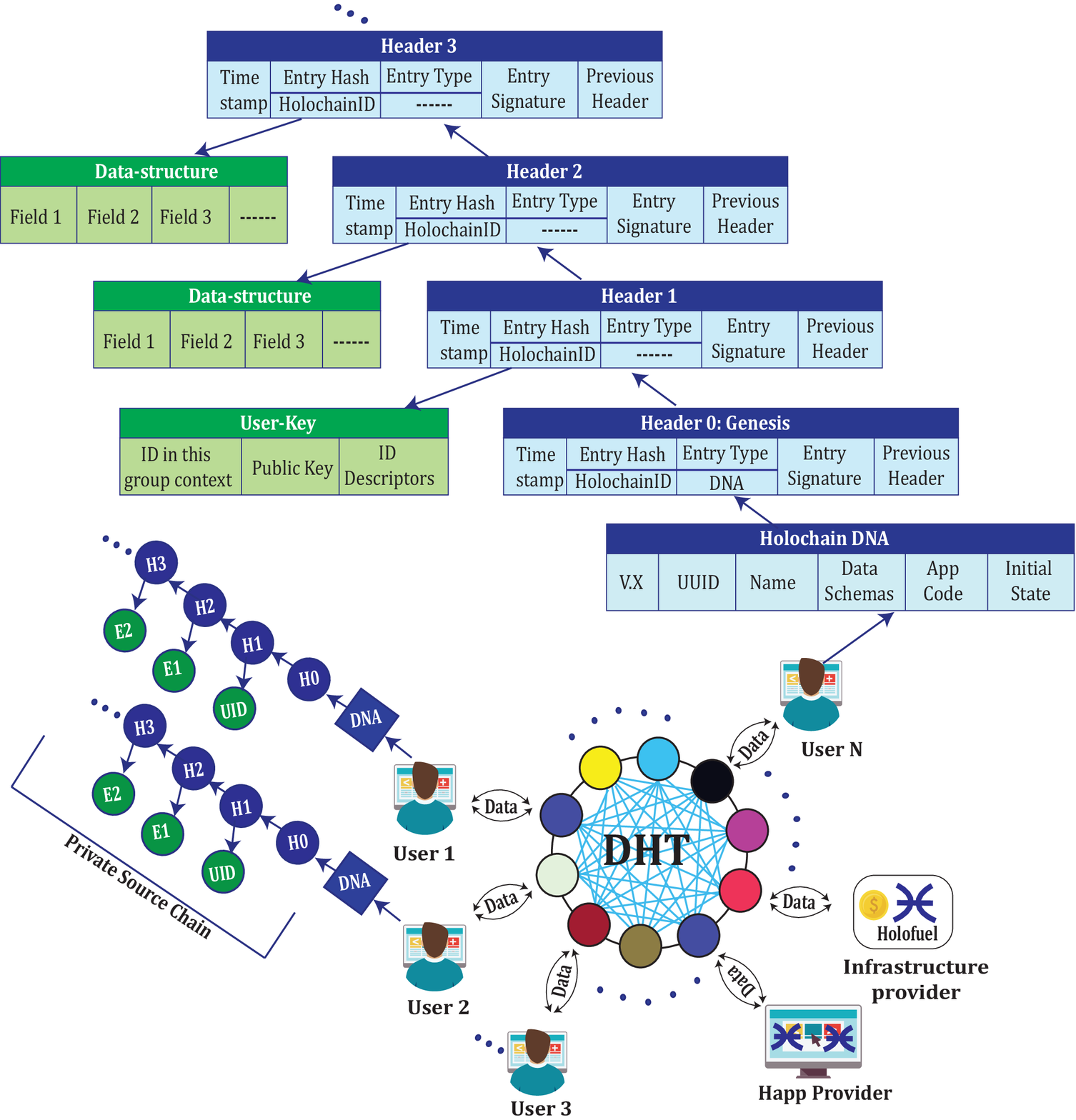}
\caption{The source chain structure of a holochain network.}
\label{fig:Layer1}       
\end{figure*}

\subsection{Hashchain}
Hash technique is used to create a unique signature for a block of holochain data which guarantees that it cannot be tampered or lost once it has been formed, propagated and stored in other peer nodes in a distributed network \cite{noauthor_redistributive}. For instance, each transaction of a holochain contains the previous header hash value that ensures the integrity of the whole source chain entries. Each block contains the previous header hash which strengthens the security of the last block while the last block points to the second last block through the hash of the previous block, and so on. Therefore, the source chain is a sequential block of all transactions that originates from the first one. Fig.~\ref{fig:Layer1} illustrates the hashchain structure in a local source chain of a user where each entity or block contains its previous header's hash value to ensure data integrity. If an attacker wants to alter one transaction block of a source chain, the previous header hash will also be changed and thereby the tampering can be easily detected by the users.

\subsection{DHT}
DHT is a distributed data storage approach that creates a hash table on an authorized P2P network. DHT was initially introduced to allow a large number of peers to transfer their confidential information locally in a holochain network. With the introduction of Bitorrent and Napster model, it has evolved to be a more efficient, powerful, and fault-tolerant technique capable of handling each node's propagation regardless of whether the node is online or offline \cite{Harris-Braun_Holochain_2019}. Each user of an hApp is connected to a DHT to share their sensitive information with each other as presented in Fig.~\ref{fig:Layer1}. For instance, patients, doctors, medical staff, and other technologists in a smart healthcare system are connected in an identical DHT for a specific hApp to share and store their information which ensures distributed data integrity.

\subsection{Source Chain Structure}
Each agent of a holochain network stores the information of every transaction together with the validation rules and applications source code locally. The structure of a whole chain, referred to as a source chain, is represented in Fig.~\ref{fig:Layer1}. The source chain of an agent holds three components: DNA, genesis and transactions.

\subsubsection{DNA}
The DNA is a set of validation rules which is unique for each hApp and applies to every user of an hApp. It ensures the integrity of the network without global consensus like blockchain. This unique feature equips holochain with scalable capabilities. The existence of different validation rules in holochain uniquely identifies different source chain addresses. A DNA file consists of multiple components such as application name, description of the holochain, the DHT structure, app specific data schemes as well as functions that execute the app operations. 

\subsubsection{Genesis}
When a user installs a new hApp, a hash is created to guarantee that the app is following the predefined rules. This hash is known as genesis. It is stored as the second entity of the local source chain. A genesis has two special entries: the first one is the hash of the DNA that applies the agreement rules to the user and the second one is the agent ID that holds the user's public key. Additionally, it can also contain information like invitation code or payments schedule. Whenever a user creates a new entity of the hashchain, it will look back to the genesis block to ensure whether the new block is following the given rules.

\subsubsection{Block of a Holochain}
Each block is stored in the local chain as an entry. Each entry contains several elements in its header such as timestamp of creating the transaction, entry type that defines the purpose of the data, hash of the data, digital signature of the entry, and hash of the previous header. Moreover, the block also contains users identity and private data of a transaction along with the header \cite{Holochain_org}.


\section{Implementation of Holochain in IoT Healthcare Systems}

Most of the healthcare facilities offer app-based services where different types of users of a healthcare system sign-up to get continuous support and monitoring of health conditions. In a smart healthcare ecosystem, holochain-based healthcare applications can be designed which work together like a marketplace to provide services. Healthcare providers offer various smart facilities by introducing a specific set of rules/protocols under which a user can benefit themselves. Users can search the whole marketplace to find their most desirable offers and accept the public agreements. Then health service delivery apps are initiated to build a private relationship between healthcare providers and the users.

The proposed holochain framework facilitates completely distributed IoT-based smart healthcare systems, which is represented in Fig.~\ref{fig:holochain} where holochain is implemented for storing information and ensuring the security and privacy of the network. Holochain can be implemented at the network edge (IoT nodes, fog nodes, etc.) as well as in cloud servers. As holochain is an agent-centric framework, each hApp of the smart healthcare entities is considered as an agent/user who can actively participate in the network to transfer the information \cite{brock_holochain_2020}. 
Moreover, we propose to use the cloud to process and store the transaction of a holochain network. Since IoT devices are resource-constrained such as limited memory, power, computation capacity and energy, it is quite inefficient and often impractical to process the complex calculations and store a large volume of data at the local storage of IoT nodes/sensors. Therefore, it is important to shift all these capability-demanding activities to the cloud servers yet facilitating the healthcare services at IoT end users.

In the proposed smart healthcare system, various stakeholders (patient, doctor, admin staff, technologist, etc.) are represented as individual agents, each of which maintains a unique source chain to store their transactions locally with the help of cloud servers as shown in Fig.~\ref{fig:holochain}. An agent can decide to use one or multiple healthcare applications depending on their own needs. Each hApp is recognized through its unique private digital signature--DNA which consists of a set of initial entry types, executable codes and parameters for the specific applications. Even though two different healthcare hApps use the same property to write the code, two different names will result in two unique DNAs.

When an agent of a healthcare system joins a hApp network, he creates an identifier by producing a key pair that consists of a private key and a public key. The key pair helps to create a unique identification of the agent into the network, ensure data authorization, access the data, and permit others to analyze and detect various types of threats and attacks. The private key is secretly stored at its own node which serves as a password and produces a digital signature using DNA, which is needed to publish as open-source resources with the public key of the agent to other peers that can also be used as an address (identifier) of the agent in the holochain network. However, with the help of the public key, other peers can authenticate the integrity of the agent’s digital signature and process encrypted data to send only for the specific user.

Instead of having a global shared consensus, all agents of a holochain have their individual local source chains to store and validate each transaction. Communications among multiple users are signed by each agent involved and are restricted to their own source chains. They are capable of transferring the health data through the identical public DHT.

Each healthcare agent maintains a secure private group of peers who share new transaction details, validity of the transaction, source of the information, the sender’s chain header containing historical sequence, peer creation and states of the network health within the group  \cite{Harris-Braun_Holochain_2019}. When a doctor wants to monitor his patient remotely using the IoT network, (s)he requests a set of health reports. The patient generates a capability grant or token for the particular reports or medical data that (s)he wants to share and stores the new transaction or stories as the new entity of the holochain. Moreover, the patient also shares the hash of the grant entry with the authorized doctor that will be used as a capability token. 

On the other hand, the doctor preserves this token as the new entry on his private source chain and uses it whenever needed to access that particular patient's data. The patient also checks the validity of the granted token and sends the required medical records to the doctor whenever needed. DHT guarantees the reliability of the distributed holochain network using gossip protocol. If a peered agent breaks the validation protocol, it will be excluded from next event participation to avoid bad-action. The agents of the holochain network use gossip protocol to share their own experience of the other agent’s behaviour. Each holochain user maintains an experience matrix which includes confidence of the experience that refers to the behaviour or outcomes of the previous experience. The confidence value of the experience matrix can be modified according to the attitude of the users. For instance, if a patient tries to double spend regarding a transactions' holo fuel value and be detected as bad-actor, then the confidence value will be decreased. High confidence agents are encouraged to validate and participate more frequently. Moreover, the value of the experience matrix can be updated by both direct experience or through other agents. A step-by-step processing of a holochain framework is presented in the Algorithm~\ref{algo1} and the channel authentication process is demonstrated in Algorithm~\ref{algo2}. Only a valid healthcare application user can create a transaction and participate in data sharing using DHT. To be a authenticated channel, a user needs to be a valid user of a specific healthcare application in the marketplace. Then any attempted transaction should be validated by the set of given rules in the DNA. 


\begin{algorithm} [!ht]
    \caption{A step-by-step implementation of the holochain framework}
  \begin{algorithmic}[1]\label{algo1}
      \STATE \textbf{
     Setup hApp consensus or validation protocol.}
      \STATE \quad  \textbf{Initialization DNA: } Design DNA as local source chain $SC_{local}$ $1^st$ entity for each ${hApp_{i}}$ in healthcare marketplace $F_{hAppi}$, where i=1, 2, 3, .....N.
      \STATE \quad \textbf{Function} create DNA in $SC_{local}$
      \STATE \quad \quad Set entity types $e_x$ as validation rules.
      \STATE \quad \quad Set executable functions $ef_x$ for specific ${hApp_{i}}$ 
      \STATE \quad \quad Set other expected parameters $p_x$  to  specified a unique ${hApp_{i}}$.
      Where, x=1, 2, 3, 4, ......N.
      \STATE \quad  \textbf{End Fuction}
      \STATE \quad \textbf{Function} Creating DNA for ${hApp_{i}}$. 
      \STATE \quad  \textbf{Initialization Genesis:} The Second entity of a local source chain $SC_{local}$.

      \STATE \quad \textbf{Function} create Genesis in $SC_{local}$.
      \STATE \quad \quad Calculate timestamp ${t_{stamp}}$ of the genesis.
      \STATE \quad \quad Initialize Private and Public key set $ID(Pr_{k}, Pb_{k})$. 
      \STATE \quad \quad Calculate hash value of the DNA.
      \STATE \quad  \textbf{End Fuction}
      \STATE \quad \textbf{Function} Creating Genesis 
      \STATE \textbf{Create temper-proof hashchain based new holochain entity.}
      \STATE  \quad \textbf{Function} Demands for a new transaction/entity.
      \STATE  \quad  \quad Calculate  ${t_{stamp}}$ of the new entity.
       \STATE  \quad  \quad State the new entry type.
       \STATE  \quad  \quad Calculate digital signature using step 26 .
        \STATE  \quad  \quad Create current data hash.
      \STATE  \quad  \quad Calculate hash of previous header ${hash_{pvh}}$.
      \STATE  \quad  \quad  Store signed entity in $SC_{local}$ before broadcast.
      \STATE  \quad \textbf{End Function}
      \STATE  \quad \textbf{Function} creating a new holochain entity.

     \STATE \textbf{Cryptographycally signed each hashchain entity. }
     \STATE \quad \textbf{For} each new entity of a holochain.
     \STATE \quad \quad Calculate digital signature of the transaction using agent's private key.
     \STATE \quad \quad Store signed entity in hashchain based $SC_{local}$ before broadcast.
     \STATE \quad \textbf{End For}

      \STATE \textbf{Design DHT to broadcast the valid transaction among peered users via gossip protocol.}
      \STATE \quad \textbf{Create DHT for authenticated channel.}
      \STATE \quad \quad  Initialize a set of ($Pb_{k}$, $hash_{entity}$). 
    \STATE \quad \quad Share ($Pb_{k}$, $hash_{entity}$) with random users who have same DNA.
   \STATE \quad \quad Validates the transaction using their own copy of DNA.
 \STATE \quad \quad Valid transaction is broadcast to other users to backup through gossip protocol.     
    \STATE \textbf{Setup gossip protocol to resist bad data broadcast.}
    \STATE \quad \textbf{Design gossip protocol.}
    \STATE \quad \quad  Create matrix $M$ of a set of $(\gamma,~\eta)_{self}$, $(\gamma,~\eta)_{others}$, where, experience $\gamma$ and confidence $\eta$ refer to the behaviors of other nodes.
    \STATE \quad \quad  Update $(\gamma,~\eta)_{self}$, $(\gamma,~\eta)_{others}$ in each experience to resist bad users entity.     
  \end{algorithmic}
\end{algorithm}

\begin{algorithm} [!ht]
    \caption{Channel validation algorithm}
  \begin{algorithmic}[1]\label{algo2}
   \STATE \textbf{
     Transaction validation of a   \textbf{$hApp_{i}$} }
     \STATE \quad \textbf{For} ($e_x=1; ~ e_x <N; ~ e_x ++$)
     \STATE \quad \quad \textbf{If} { (all $e_x \in$ DNA)} \\
     \STATE \quad \quad {Ensure the validity of the transaction}
      \STATE \quad \quad \textbf{Else}
     \STATE \quad \quad {Invalid transaction}
     \STATE \quad \quad \textbf{End If}
     \STATE \quad \textbf{End For}

      \STATE \textbf{
     Application validation of a   healthcare marketplace. }
     \STATE \quad \textbf{For} (all $e_x$ $\in$ DNA)
     \STATE \quad \quad \textbf{If} { ($hApp_{i}$ $\in$ $F_{hApp}$)} \\
     \STATE \quad \quad {Ensure the validity of the $hApp_{i}$. }
      \STATE \quad \quad \textbf{Else}
     \STATE \quad \quad {Invalid application.}
     \STATE \quad \quad \textbf{End If}
     \STATE \quad \textbf{End For}
   
   \STATE \textbf{ 
   Overall channel authentication  }  
    \STATE \quad \quad \textbf{If} { Validation in (Step $1$ $\&$ Step $7$ ) succeeds,} \\
     \STATE \quad \quad {Channel authentication complete.}
      \STATE \quad \quad \textbf{Else}
     \STATE \quad \quad {Invalid channel.} 
     \STATE \quad \quad \textbf{End If}
     
  \end{algorithmic}
\end{algorithm}

\section{Security Analysis}

\subsection{Privacy and Security}
DLT records transaction details, replicates, synchronizes and transfers digital information across all over the network in a distributed fashion. Holochain is a security-preserving DLT technology as it implements the concept of both advanced cryptography and cryptocurrency (holo fuel). Holchain is reliable, tamper-proof and resistant to various attacks such as Denial-of-service (DoS), fake node, Man-in-the middle attack (MitM), double spending and illegal data tempering.

\subsection{Security Threats}
Smart healthcare services deal with a large number of significant sensitive personal data of the users. Moreover, due to the heterogeneous technologies of an IoT-healthcare system, security vulnerabilities are needed to be considered. Some frequent threats or attacks are discussed below.

\subsubsection{Unauthorized Access}
Unauthorized access occurs when an intruder wants to access healthcare information in a network through compromised nodes or communication channel without appropriate authorization or permission. The malicious user sometimes manipulate, alter, destroy or gain ownership of confidential health information using unauthorized access. The works in \cite{Unauthorized_2018} analyzed the vulnerabilities of using simple and default password in a healthcare IoT environment. It demonstrated that through unauthorized attack using default Secure Shell (SSH) commands, it was able to launch brute force attack and gained access to the IoT nodes (Raspberry Pi) for modifying and forging crucial personal data.
%


\subsubsection{Illegal or Intentional Data Tempering}
Illegal data tempering is one of the frequent attacks on IoT healthcare networks. It can cause data integrity issues. The attacker of this threat could be an insider who can temper its own node information in a holochain or even an outsider. The work in \cite{liu_enhancing_2020} demonstrated how an illegal data tampering attack on biomedical security systems can be done to breach healthcare records from the communicating nodes to track the communication or alter the actual data.


\subsubsection{MitM}
MitM occurs when an attacker sends a malicious request or eavesdrop in order to to monitor, access and modify a transaction between two agents of a holochain network. The works in \cite{noauthor_ic-mads_nodate} demonstrated a cross-layer MitM attack in a smart healthcare application. The performance evaluation shows that with the increasing number of attackers (from 0 to 25), communication overhead is increased to 15\% from 10\% while the packet delivery ratio is  dropped to 69\% from 96\%.


\subsubsection{DoS/Distributed DoS (DDoS)}
Each layer of an IoT network may be compromised by DoS/DDoS attack in both the IoT nodes and network links. DoS/DDoS occurs when an attacker sends a malicious request flood to disrupt the functionalities of targeted IoT nodes or cloud servers or communication links. This attack is also responsible for making dedicated services unavailable for an authorized user. The authors in \cite{kumar_enhanced_2020} analyzed the weaknesses of the \emph{Datagram Transport Layer Security} (DTLS) protocol which is employed in a constrained healthcare network to preserve the security of health data. DTLS can be compromised by a large number of ClientHello messages sent by an attacker to create a DoS attack for establishing fake communication between attacker and server. This will occupy legitimate bandwidth and resources for each ClientHello message.

\subsubsection{Double Spending Attack}
In holochain networks, an agent can duplicate or reuse the digital token or cryptocurrency (i.e., using the same holo fuel token multiple times) and transmit as identical tokens to multiple receiver agents. Double spending problem is a serious security threat for the various smart applications including healthcare. In  \cite{cullen_distributed_2019}, a channel parasite attack was implemented to design double-spending attack on a blockchain-based IoT framework which disrupted the immutability and irreversibility of the DLT.


\subsection{Protection against Threats}
Here, we discuss how holochain addresses the above threats.

\subsubsection{Unauthorized Access}
The DNA limits the unauthorized user access of a holochain network. Every user of a network should have a unique DNA that provides capability-based security and ensures access control to the users as well as its source chain data \cite{Holochain_developer}. 

\subsubsection{Illegal Data Tempering}
If a user modifies his own code of an hApp, it will automatically redirect the hApp services to a completely different hApp network from the authorized shared DHT network. Therefore, if a user wants to modify the code intentionally, he cannot influence, access or modify the records of the original network. 

\subsubsection{MitM Attack}
When an attacker initiates MitM attacks in a holochain network, it can be detected on a source chain entry through the digital signature. Each modification of an hApp creates an entry that is signed by the private key of the user and adds to the header. Therefore, the digital signature can ensure the data origins of a request. The hash value of the previous header also helps to detect MitM attacks among the entire stored entries of a source chain. Whenever an intruder tries to modify the previous entry, the hash value of the previous header will notify nodes about the unauthorized activity. 

\subsubsection{DoS/DDoS}
IoT nodes are in greater risk of getting compromised due to various types of devices with wide ranging characteristics. Holochain networks are capable of handling on-demand P2P communication against Dos/DDoS attacks. When a transaction is faced with a DDoS attack, it could demand to impose filtering rules (e.g., Border Gateway Protocol (BGP)). For instance, the transaction will have the detailed information such as type of the attack, total counted drop packet and average dropped packets per unit time which will be signed to store in local hash-chain and will be broadcast through DHT. After broadcasting the transaction, the service provider can employ a DDoS attack detection method by analyzing the record samples with the help of BGP to validate whether DDoS is initiated on a targeted node \cite{holo_ddos}.
 
\subsubsection{Malicious Code/Node}
When an attacker attempts to hack the holochain code, random peers will fail to validate the generated anomalous results. Therefore, the stored abnormal outputs will be tagged as counterfeit and will not be transmitted. The network is thus capable of identifying the bad transaction and can blacklist the agent who commits the crime.


\subsubsection{Double Spending Attack}
Since the holochain technology uses electronic credit currency to communicate with other users in the network, ensuring  transparency can handle the double spending attack robustly. For instance, A is the patient of a holochain-based system who has 5 holo fuel credit to make a creditable transaction that sends to a doctor X. Assume that after sending the credit to the doctor X, patient A intentionally removes the A-X transaction and again sends the same balance to another doctor Y for his service. The holochain network will detect this kind of double spending attack through gossip protocol. Whenever patient A makes A-X transaction, gossip protocol will be responsible for spreading the news of the occurred transaction to some randomly selected users. When patient A tries to double spend with doctor Y, the network checks with those randomly selected users' log of the transaction history to validate whether the balance is correct.

\section{Performance Analysis of Security Mechanisms} 

In this section, some prevalent lightweight cryptographic algorithms are considered with their performance evaluated and compared using medical data in IoT networks against two key performance indicators (KPI), namely memory usages and CPU cycles per bits. A comparative analysis on how various existing DLTs perform against the KPIs are also provided. Among them, AES and Data Encryption Standard (DES) are two popular symmetric key block cipher algorithms while RSA is used as an asymmetric key encryption mechanism. They are able to detect and resist various common attacks in IoT such as MitM attack \cite{Springer_misc}. Leak Extraction (LEX) and Light Encryption Device (LED) are extended versions of AES. In addition, LEX is a software-oriented stream cipher which modifies the AES key stream using a recursion process. On the other hand, LED is more useful for hardware implementation which uses a simple key schedule to resist various attacks. This is particularly true for LED-80 \cite{Springer_misc}.

\begin{figure}

  \includegraphics[width=0.5\textwidth]{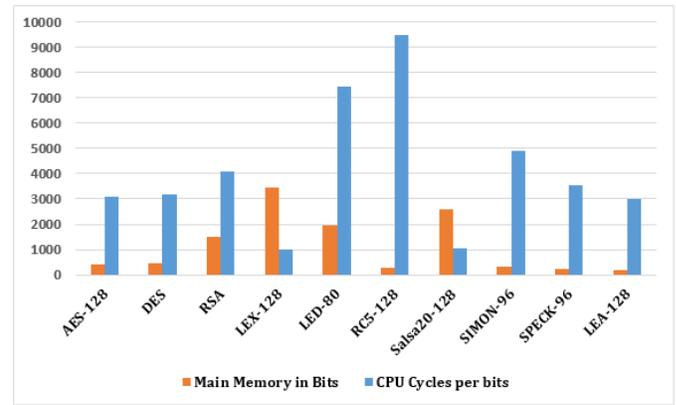}
\caption{Comparative analysis of existing encryption mechanisms in IoT networks.}
\label{fig:security}       
\end{figure}

Another important type of cryptographic algorithms is categorized as lightweight block ciphers such as RC5 and Salsa20 \cite{jrnl_lwc_survey}. RC5 utilizes a variable number of block size, key size and number of rounds during recursion which depends on the functionality of microprocessors.  Salsa20, on the other hand, utilizes the advantage of hash and XOR functions with a 64-byte block size \cite{guneysu_lightweight_2016,meiser_efficient_2008}. Both RC5 and Salsa20 are more appropriate for IoT-based medical applications due to the relatively lower memory requirements. However, SPECK and SIMON are used for multi-block cipher with a variable number of key size and block size. The fundamental advantage of using SIMON and SPECK is improving the speed and memory utilization which is more suitable for a lightweight healthcare application. Fig.~\ref{fig:security} shows a comparative performance analysis of existing IoT security mechanisms as presented in \cite{jrnl_lwc_survey}. Fig.~\ref{fig:security} also demonstrates that compared to the other considered ciphers, LEX is the fastest in terms of CPU speed. Though AES and DES are slightly faster than SPECK, in terms of memory usages, SPECK is faster than AES and DES.  Considering all the various security mechanisms' performance, it is suggested that SIMON and SPECK offer better performance in resource-constrained IoT networks.

\begin{figure}

  \includegraphics[width=0.5\textwidth]{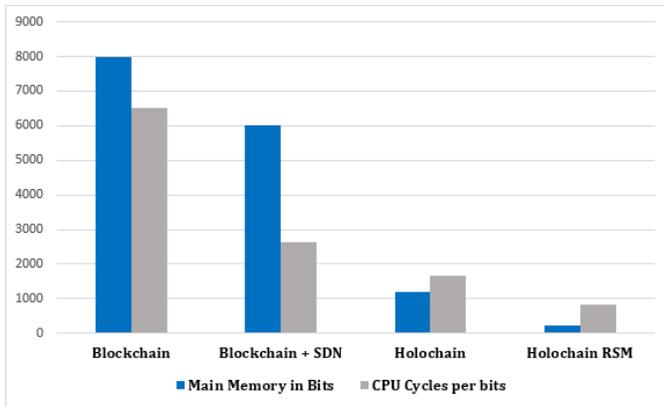}
\caption{Comparative analysis of blockchain and holochain based encryption mechanisms in IoT networks. }
\label{fig:security-bchc}       
\end{figure}

On the other hand, Fig.~\ref {fig:security-bchc} includes the comparative analysis of the performance of popular DLTs in terms of IoT security. Since the functionalities of DLTs are different from the existing traditional cryptography mechanisms, the memory requirement more important than the CPU cycle time. According to \cite{Dinh_BC_2018}, the hybrid technology of Software-Defined Network (SDN) and blockchain provides better performance compared to traditional blockchains. Blockchain includes the request from any user, but SDN ensures the secure connection and avoids unnecessary requests, which reduces the memory as well as CPU cycle per bits. Though this technique brings new breakthroughs in the world of blockchain, the memory requirements and processing techniques remain the challenges. However, holochain and the new version of holochain (Holochain RSM) reduce huge loads of data processing and storing in the dynamic and real-time implementation like IoT networks. The memory utilization and speed of holochain are far better than that of the blockchain technology which can be an out of bound thinking for the upcoming IoT networks.


\section{Future Directions}\label{sec_future}
The holochain-based technique will play a significant role to ensure security and privacy for next-generation communication models for large-scale deployment as it allows the advantages of high scalability, lightweight and decentralized architecture, flexibility and transparency with a high level of security. The characteristics of being lightweight, transparent and distributed operation ensures fast processing which is crucial for high data-rate and low-latency communication systems emerging in 5G or in upcoming 6G standards. Moreover, the holochain technology will also be utilized for storing massive data in a distributed fashion with the help of fog nodes, cloud servers, mobile edges, and etc. Undoubtedly, holochain has a significant impact on distributed security but more advances need to be made to suit the resource-constrained IoT environments for reducing latency of a transaction, processing speed, and tackling real-time threats. To meet those needs, there are a number of challenges that need to be addressed.



\subsection{Real-Time Cryptocurrency Processing and Monitoring}
In recent computing systems, cryptocurrency emerges as an alternative to the physical currency that has been growing. However, processing the real-time holo fuel and analyzing the performance in IoT networks can be a difficult task. As  mentioned earlier, thousands of transactions are in need of processing and transferring among various agents through the holochain. Therefore, cryptocurrency is required to support the network which requires the presence of a unique real-time monitoring system used by all agents of holochain networks.

\subsection{Real-Time Smart Threats Detection}
With the advances in technology, an attacker may create ML-based intelligent and unknown attacks on IoT nodes in a  holochain-based smart healthcare system. Implementation of a complex and intelligent real-time threat detection model requires a high degree of computation capacity and memory which is quite challenging for resource-constrained IoT nodes. Therefore, it is imperative to design a real-time smart threat detection mechanism suitable for running on memory and processing capacity constrained systems.

\subsection{Load Balancing}
With the increasing number of users, a holochain network distributes its load over other authorized peers. The peers are selected using specially designed protocols to randomly select nodes in the network. It needs to ensure that alternate nodes are available with validation power or storing capability of other users information whenever one goes off line. Finding the right balance to select the minimum number of agents while ensuring the process of validation is intact is a challenging task which becomes more difficult with an  increasing scale of the network size. Moreover, this whole process of assigning and reassigning agents should be near real time to avoid any disruption in the operation of the IoT healthcare system.

\subsection{Quick Response to Bad Activity} 
In holochain networks, information storage, access and sharing authority are not always fixed. For instance, if one patient loses awareness, (s)he can no longer possess the authority of sensitive health related data sharing, but still, this patient can be the best choice candidate for sharing the data. Thus, quick response to bad-activity and detecting the responsible agents are essential to maintain a healthy holochain.

\section{Conclusions}\label{sec_con}
A holochain-based privacy-preserving secure communication scheme for distributed IoT healthcare applications has been proposed in this article which leverages the inherent autonomy of the holochain architecture and protocols. In contrast to blockchain, holochain liberates the communicating agents from any form of centralized control by running the applications (hApps) entirely at the user side. Therefore, there exists no central point of failure. Since users are the hosts, as more agents use an app, more hosting power and storage become available and the load gets lighter. If any agent alters their own app code, they effectively fork themselves out of the shared DHT space into an entirely different application. Thus holochain has appeared to be the most effective technology for distributed IoT applications. Comparative performance results and analyses demonstrate significant reduction in time and space complexity of the holochain framework compared to the rival blockchain schemes, which shows promises for realistic deployment of large-scale IoT healtchare systems.

\bibliographystyle{IEEEtran}\footnotesize{
\bibliography{IEEEabrv,ref1.bib}}

\end{document}